  \documentstyle[12pt]{article}
\newcommand{\bmth}[1]{\mbox{\boldmath{$ #1$}}}
\newcommand{\epsl}{\varepsilon}

\title{
 \begin{flushright} {\normalsize KNK 962} \end{flushright}
 \vspace{1.0\baselineskip}
 Two-Body Dirac Equation and Its Wave Function at the Origin
 \thanks{A preliminary report was presented in the workshop 
"Fundamental Problems in the Elementary Particle Theory" held at 
Nihon University in March 1995. }
}
\author{{
Hitoshi I{\sc to}} \\
 \normalsize \it Department of Physics, Faculty of Science and
Engineering \\ \normalsize \it Kinki University, Higashi-Osaka, 577 
J{\sc apan}}
  \date{\normalsize \it December 18, 1996 }
\begin{document}

\maketitle
 \subsection*{Abstract }
 \small

A relativistic equation is deduced for the bound state of two 
particles, by assuming a proper boundary condition for the 
propagation of the negative-energy states. It reduces to the 
(one-body)Dirac equation in the infinite limit of one of
 the constituent mass. It also has the symmetries to assure the
 existence of the anti-bound-state with the same mass.
The interaction kernel(pseudo-potential) is systematically
constructed by diagonalizing the Hamiltonian of the background field
 theory, by
 which the retardation effects are included in the interaction. 
Its wave function at the origin(WFO) behaves
 properly in a manner suggested by the covariant field theory.
 The equation can well be applied in the heavy quark effective 
theory.

\normalsize

\vspace{10mm}
Pacs: 03.65.Pm, 11.10.St

{\bf Keywords}

 two-body equation \hspace{10mm} Dirac equation \hspace{10mm}
 relativistic dynamics \hspace{10mm} 

  wave function at the origin
 \hspace{10mm} $PCT$ invariance  \hspace{10mm} heavy quark


\section{Introduction}

One of the unsatisfactory nature of the Bethe-Salpeter equation for 
the bound state is that it does not reduce to the Dirac equation in 
the infinite limit of the one of the constituent mass, even when the 
interaction is assumed to be
 instantaneous\cite{Salpeter52}. We have to sum up all the
crossed diagrams to recover the Dirac equation\cite{Brodsky69}, which
 is impossible for the finite masses. The reducibility to the Dirac 
equation is essential in applications to the heavy flavored quarkonia 
or the other atom-like systems.

Historically,
 the relativistic single-time equation for the two-body system 
preceded the BS equation. Soon after the discovery of the Dirac 
equation, Breit proposed the equation of the form\cite{Breit29}

\begin{equation}
  \{ H_1(\bmth{p}_1)+H_2(\bmth{p}_2)+V \}\psi=E\psi,  \label{1}
\end{equation}
where $H_i$ is the Dirac Hamiltonian

\[  H_i(\bmth{p}_i)= \bmth{\alpha}_i\cdot\bmth{p}_i + m_i\beta_i   \]
and $V$ is a local potential. The Breit equation reduces to the Dirac
 equation in the limit mentioned above but does not have the 
``$E$-parity symmetry", by which we mean there is symmetric
 negative eigenvalue $E$ for each positive
one, which is interpreted as the bound state of the antiparticles.

$E$-parity symmetry is a result of the $PCT$ invariance.
 The Dirac Hamiltonian is odd under the $PCT$ transformation and the
interaction Hamiltonian considered below is invariant under it. We 
therefore see that the $E$-parity symmetry in the instantaneous BS
 equation is assured by a projection factor
\begin{equation}
     \Lambda_{++}-\Lambda_{--}    \label{2}
\end{equation}
 in front of the potential $V$, which consists of the
 energy-projection operators

\begin{equation}
  \Lambda_{\epsl\eta}(\bmth{p}_1,\bmth{p}_2)=
\Lambda_\epsl^1(\bmth{p}_1)
  \Lambda_\eta^2(\bmth{p}_2), \hspace{5mm} \epsl, \eta=+\mbox{ or }-,
 \label{3}
\end{equation}
where

\[ \Lambda_\epsl^i(\bmth{p}_i)=\{E_i(p_i)+
\epsl H_i(\bmth{p}_i)\}/2E_i(p_i), \]
\[ E_i(p_i)=(\bmth{p}_i^2+m_i^2)^{1/2}.  \]
We must introduce a similar factor in any attempt at the
construction of an improved two-body equation.

The factor (\ref{2})
 comes from the St\"{u}ckelberg-Feynman boundary condition
 for the propagation of the negative-energy
 state\cite{Stuckelberg41}.
 We will construct the equation for the unequal-mass constituents by
imposing similar boundary condition.
 Before presenting it, we restrict the framework of consideration. 
For definiteness, we assume Abelian gauge field interacting with the
 Dirac particles. We also work in the rest
 frame of the bound system($P=(E,\bmth{0})$),
 since we are looking for the non-covariant
approximation of the low-energy dynamics. $\bmth{p}$ and $\bmth{x}$,
 in the
 following, are the relative momentum and coordinate, respectively,
 in this frame.

\section{The equation for unequal-mass constituents}

In this section, we deduce a new single-time equation(Two-body Dirac 
equation) for the unequal-mass constituents, for which we assume that 
the mass $M(m_1)$ is larger than $m(m_2)$.

\subsection{\it Two-body Dirac equation}

Let us first consider the 2-body propagator for which
 we impose the boundary condition that the negative-energy states
propagate backward in time. If we keep individuality of the 
constituents and use the usual Feynman propagator in the instantaneous
 BS equation, the equation (\ref{1}) modified by the factor 
(\ref{2}) results.
However, we can choose the other possibility in which
 we incorporate the idea that the bound two bodies should
 be treated as a quantum-mechanical unity.\footnote{We can establish
 the concept of individuality in the quantum
mechanics only through observation, which brings about a subtle point
 to the bound system:
 To detect an individual one in the bound constituents, we
need to separate them by applying the 3rd interaction, which 
inevitably destroys the original state. So there is no {\it a priori}
 reason to apply the free propagator individually to each constituent.
 Note that ``the free propagator'' itself {\it for the bound state} 
is merely a convention of approximation. }
 Since $M$ is larger than $m$,
 the free part of the Hamiltonian has the same sign as that of the 
particle 1 in the CM system. Let us then  modify the boundary 
condition as follows:
{\it A bound two-particle state propagates backward in time if
 their net energy is negative.}  By assuming it, we have the propagator

\begin{equation}
  S_F^{(2)}(P,p)=\sum_{\varepsilon\eta}
\frac{1}{ \lambda_1 E-p_0-H_1(-\bmth{p})+i\varepsilon\delta}
\frac{1}{\lambda_2 E+p_0-H_2(\bmth{p})+i\varepsilon\delta}
\Lambda_{\varepsilon\eta}\gamma_0^1\gamma_0^2,    \label{2p}
\end{equation}
where $\lambda_1+\lambda_2 =1$ and the limit $\delta\rightarrow +0$ 
is assumed.

By assuming (\ref{2p}) we obtain the single-time equation. The
 pseudo-4-dimensional form of it, in the momentum space, is

\begin{equation}
 \psi(p)=iS_F^{(2)}(P,p)\int V(\bmth{p},\bmth{q})\psi(q)d^4q/
                                (2\pi)^4,     \label{p4}
\end{equation}
where $V(\bmth{p},\bmth{q})$ represents the interaction, which does
 not depend on the relative energies but is not necessarily an
 instanteneous local potential. 
After integrating out the redundant degree of the freedom, 
(\ref{p4}) becomes

\begin{equation}
  \{ H_1(-\bmth{p})+H_2(\bmth{p})+\sum_{\epsl\eta}\epsl
   \Lambda_{\epsl\eta}V \}\psi=E\psi,  \label{4}
\end{equation}
which reduces to the Dirac equation in the infinite limit of $M$.
 It is easy to see the $E$-parity symmetry of this equation.

If we would apply our equation to the scattering state, we shall
 have, from the time-dependent equation, the conserved probability 
density

\begin{equation}
   \rho(\bmth{x},t)=\psi(\bmth{x},t)^\dagger \sum_{\epsl\eta}\epsl
     \Lambda_{\epsl\eta} \psi(\bmth{x},t),   \label{5}
\end{equation}
which is in accord with the boundary condition that the negative-
energy state propagates backward in time carrying the negative 
probability density\cite{Aitchison82}.
However, it does not necessarily provide the normalization condition for
 the bound state: For the scattering processes, the projected wave 
function $\Lambda_{\epsl\eta}\psi$ corresponds to the physical state
 of the (free)particles with the positive or negative energy $E$. And
 the above interpretation of the probability current actually says
 that {\it the state with the negative $E$ is carrying the negative
probability density}. However, for the
bound state with the positive eigenvalue $E$,
 $\Lambda_{--}\psi$ or $\Lambda_{-+}\psi$ is merely
 {\it a negative-energy component in the representation in which the
 energy of the free particle
 is diagonal}.

Taking the above consideration into account, we restore the
 probability interpretation of the wave function and normalize it by
 assuming the probability density

\begin{equation}
   \rho(\bmth{x})=\psi(\bmth{x},t)^\dagger \psi(\bmth{x},t). 
             \label{6}
\end{equation}
Observables except the Hamiltonian are self-adjoint under this
 metric:
\begin{equation}
  (\phi,\hat{O}\psi)=(\hat{O}\phi,\psi).                \label{66}
\end{equation}
The Hamiltonian is the operator ruling the time development of the 
system and its interaction part is modified by the factor 
$\sum_{\epsl\eta}\epsl\Lambda_{\epsl\eta}$. Though
 it is not a self-adjoint operator, its eigenvalue is proved to be
 real if the inner product (\ref{g3}) below exists.

 \subsection{\it Green's function and the vertex equation}

We define the Green's function $G$ for (\ref{4}) by the operator 
equation

\begin{equation}
  \{E- H_1 -H_2 -\sum_{\epsl\eta}\epsl\Lambda_{\epsl\eta}V \}G
          =\sum_{\epsl\eta}\epsl\Lambda_{\epsl\eta},  \label{g1}
\end{equation}
and
\begin{equation}
  G\{E -H_1 -H_2 -V\sum_{\epsl\eta}\epsl\Lambda_{\epsl\eta} \}
          =\sum_{\epsl\eta}\epsl\Lambda_{\epsl\eta}.  \label{g2}
\end{equation}
In the momentum representation, it can be written, by using the 
eigen-function
$\chi_n(\bmth{p})$ of (\ref{4}), as
\begin{equation}
 G(\bmth{p},\bmth{p}')=\sum_n \frac{1}{N_n(E-E_n)}
      \chi_n(\bmth{p})\chi_n(\bmth{p}')^{\dag} +\mbox{continuum},
\end{equation}
where $E_n$ is an eigenvalue and $N_n$ is a normalization factor
 defined by
\begin{equation}
  N_n=(\chi_n,\sum_{\epsl\eta}\epsl\Lambda_{\epsl\eta}\chi_n). 
 \label{g3}
\end{equation}
When one of the constituents(labeled with 2) belongs to the same class
 of the antiparticle of the other, there can be an annihilation process
 for which the unamputated-decay-vertex $\Phi$ is given by
\begin{equation}
   \Phi= C\gamma G,  \label{g4}
\end{equation}
where $\gamma$ is the lowest vertex and $C$ is the charge-conjugation
 matrix of the particle 2.

$\Phi$ satisfies the vertex equation
\begin{equation}
 \Phi(E-H_1-H_2- V\sum_{\epsl\eta}\epsl\Lambda_{\epsl\eta})=
         C\gamma\sum_{\epsl\eta}\epsl\Lambda_{\epsl\eta}  \label{g5}
\end{equation}
and the amputated vertex is
\begin{equation}
  \Gamma =\gamma +C\Phi V.                             \label{g6}
\end{equation}
We can determine the renormalization constant $Z_1$ for the wave 
function at the origin(WFO) from (\ref{g5}) and (\ref{g6}), if we 
need it\cite{Ito82}.

\subsection{\it Interaction Hamiltonian}

We have, so far, not specified the interaction Hamiltonian
(quasipotential).
In this section, we investigate it for the one-(Abelian)gauge-boson
 exchange in the Coulomb gauge as an example.
 For the instantaneous part of the interaction, $V$ is obvious. For
 the remaining part, we can specify the quasipotential in a clear way
 from the background field theory. We have already
 shown, for the Salpeter equation, that the quasipotential from the
 one-boson exchange is given through the diagonalization of Fukuda,
 Sawada, Taketani\cite{Fukuda54} and Okubo\cite{Okubo54}
(FSTO)\cite{Ito90}. We also apply this method to the present equation.

We introduce the generalized Fock subspace of the
 free constituents, the bases of which are denoted by 
$ |\epsl,\eta,\bmth{p}\rangle, $
where $\epsl$ and $\eta$ are the signs of the energies of the
 particles 1 and 2 respectively.\footnote{There was some error 
concerning the Fock space in Ref.\cite{Ito90}.
 Namely, we employed the usual(positive-energy) Fock space and 
reinterpreted
 the matrix elements of the interaction Hamiltonian including the
 negative-energy indices as the ones in this space according to the 
hole theory. It, however, brings the procedure into confusion, 
since we have revived the negative energy in Eq.(\ref{4}).
 However, the error is only conceptual for the Salpeter equation which
 has only projection factors $\Lambda_{++}$ and $\Lambda_{--}$.
 The result of Ref.\cite{Ito90} is correct.}
 We then diagonalize the Hamiltonian
 in the Schr\"{o}dinger picture by using the FSTO method. 
The second-order boson-exchange potential in this subspace is given
 by

\begin{eqnarray}
\lefteqn{\langle\epsl,\eta,\bmth{p}|V(\mbox{1b})|\epsl',\eta',
\bmth{p}'\rangle=\frac{g^2}{(2\pi)^3}\sum_{ij}
\alpha_{1i}(\delta_{ij}-\frac{1}{\bmth{q}^2}q_iq_j)\alpha_{2j} }
 \nonumber
  \\ & & \times\frac{1}{2} [ \frac{1}{\bmth{q}^2-
\{ \epsl E_1(p)-\epsl' E_1(p') \}^2}
+\frac{1}{\bmth{q}^2- \{ \eta E_2(p)-\eta' E_2(p') \}^2} ],
        \label{7}
 \end{eqnarray}
where $\bmth{q}= \bmth{p}-\bmth{p}'$. The retardation effects are
included in this equation.

\section{Some properties}

\subsection{\it WFO of the $\,^1S_0$ state}

Let us next study some basic feature of the equation. When it
 is applied to the system in which the pair
 annihilation of the constituents can occur, an important physical
 quantity is the wave function at the origin(WFO). For example, the
 decay amplitude of the pseudo-scalar $Q\bar{q}$ meson via a weak 
boson is proportional to the average WFO
 $\mbox{Tr}\{ \gamma_5\gamma_0\psi(0) \}$, where $\psi(0)$
is the charge conjugated(with respect to the particle 2($\bar{q}$))
 WFO. We investigate, in Appendix, the asymptotic behavior of the
 momentum-space wave function by using the method given in 
Ref.\cite{Ito82}. We assume instantaneous exchange of a gauge 
boson\footnote{The analysis in the Appendix
cannot be applied to the retarded interaction.}.
The average WFO thus obtained is finite. This result is
consistent with consideration on the covariant field theory. We note
 that the average WFO becomes divergent in the limit of the one-body
 Dirac equation,
 for which we have the renormalization procedure\cite{Ito872}.

There are many ``two-body Dirac equation'' proposed. An interesting 
one from the point of view of the present paper is the one by 
Mandelzweig and Wallace\cite{Mandelzweig87}. 
They intended to include the effects of the higher-order interaction
(the crossed Feynman diagram)
 and got an equation which has the proper one-body limit and the
$E$-parity symmetry. An important difference
 from our equation is in the average WFO considered above. It is
 divergent in their equation if the transverse part of
 the gauge-boson exchange is included\cite{Tiemeijer93}.

\subsection{\it The crossed diagram}

By modifiing the boundary condition in (\ref{4}), we have included 
some parts of the crossed diagram. To estimate this effect, we take
 up the 4th-order scattering amplitude in the Coulomb potential for 
the mass-shell particles. We are interested in the contribution of 
the intermediate state with the $\Lambda_{+-}$ projection. Now, let 
us compare the two amplitudes restricted to this intermediate state:
 the `ladder diagram'($l$) with the propagator (\ref{2p}) and the 
crossed diagram($c$) with the usual Feynman propagator. 
We see that they are different by the following factors:

\begin{equation}
  I_l=\frac{\sqrt{(\bmth{p}_2+\bmth{q})^2+ m^2} 
  -\bmth{\alpha}^{(2)}
 \cdot (\bmth{p}_2+\bmth{q})-\beta^{(2)}m }
 {\sqrt{(\bmth{p}_2+\bmth{q})^2+ m^2}
( A +\sqrt{\bmth{p}_2{}^2+ m^2}
 +\sqrt{(\bmth{p}_2+\bmth{q})^2+ m^2} }     \label{Il}
\end{equation}
for the `ladder diagram' and

\begin{equation}
  I_c=\frac{\sqrt{(\bmth{p}'_2 -\bmth{q})^2+ m^2} 
  -\bmth{\alpha}^{(2)}
 \cdot (\bmth{p}'_2 -\bmth{q})-\beta^{(2)}m }
 {\sqrt{(\bmth{p}'_2 -\bmth{q})^2+ m^2}
( -A +\sqrt{\bmth{p}'_2{}^2+ m^2}
 +\sqrt{(\bmth{p}'_2 -\bmth{q})^2+ m^2} }     \label{Ic}
\end{equation}
for the crossed diagram, 
where $\bmth{p}_2$ and $\bmth{p}'_2$ are the initial and final 
momentum of the lighter particle respectively 
and $\bmth{q}$ is the loop momentum to be integrated. 
The common element $A$ is given by
\[ A= \sqrt{\bmth{p}_1{}^2+ M^2} -\sqrt{(\bmth{p}_1-\bmth{q})^2+ M^2}.
  \]
We see that $I_l$ and 
$I_c$ have the same signature and the order of magnitude for 
momenta much smaller than $M$. 

One would intend to proceed to the next order approximation in the 
interaction $V$. He will calculate the 4th-order crossed diagram 
by assuming ordinary Feynman propagator. He will, further, have to 
add counter correction terms to compensate the part of the crossed 
diagram already included effectively in the iteration of the lowest 
kernel $V$, an example of which is given by the term with the factor 
$I_l$.

\subsection{\it On application to the heavy quark effective theory}

One of the research fields in which we can utilize the two-body Dirac 
equation is the physics of the heavy flavored quarkonia. 
The recent trends in this field 
are led by the heavy quark effective theory(HQET). The $Q\bar{q}$ 
system is well described by using the two-body Dirac equation with 
some phenomenological (pseudo)potential. If we take the heavy quark 
limit($M\to \infty$) of the quark $Q$, the equation itself becomes 
the one-body Dirac equation, which is the basic equation of HQET. 
There are, however, divergent(as $M\to\infty$) portions in the WFO, 
some average of which determines the 
annihilation decay rate of a pseudoscalar quarkonium. The matrix 
element of the current for this decay is given by

\begin{equation}
    <0|\bar{q}\gamma_\nu\gamma_5Q|Q\bar{q}>\sim C(\mu)
                              \mbox{tr}\{\gamma_\nu\gamma_5\Psi(0)\}, 
                                          \label{hf1}
\end{equation}
where $\Psi(0)$ is the WFO of the bound state and we have separated 
the correction $C(\mu)$ coming from the range of momenta $\mu\sim M$ 
in the integral over the loop momenta.
$C(\mu)$ is usually determined from the perturbative loop 
corrections summed up by using the renormalization group equation
\cite{Voloshin87}\cite{Politzer88}.

\begin{equation}
  C(\mu)=(\frac{\alpha_s(M)}{\alpha_s(\mu)})^d,\quad d=-6/(33-2N_f).
                                          \label{hf2}
\end{equation}

We can calculate $1/M$ corrections to HQET by using the two-body 
Dirac equation.

\subsection{\it Another approach to the heavy quark phenomenology}

An alternative way of investigating the heavy quark systems is due to 
direct application of the two-body Dirac equation. The binding 
interaction, in this formulation, consists of the Coulomb and the 
confining potentials. We further add the gluonic correction term to 
the potential, for which we should attach the high momentum 
cutoff\cite{Ito84}. The correction factor $C(\mu)$ depends on this 
cutoff. If we take the cutoff of the QCD scale, we should include 
all the corrections except for the Coulomb contribution to the vertex
 correction. By subtracting it in the equation (\ref{hf2}), we get
 $d=2/(33-2N_f)$ for the exponent\cite{Ito92}.\footnote{
                                              We have not assumed the 
counter correction which compensates the modification of the 
two-body propagator, since it represents some real effect of the 
crossed diagram. 
In QED for which the perturbative corrections work very 
well, we have to add the counter corrections to the vertex part.}
 In the phenomenology of the $c\bar{c}$ 
and $b\bar{b}$ systems\cite{Ito90}, we took fairly large value for 
this cutoff. If we choose the same one we only need the correction 
from the self-energy of the quarks. Then, the exponent becomes
 $d=-2/(33-2N_f)$.

\subsection{\it On the equal-mass limit}

The unequal-mass equation (\ref{4}) is well applied to the system 
with $M\gg m$. For $M\simeq m$, the reason justifiing (\ref{2p}) 
becomes obscure and we will have two different equations in the 
limit $M=m$. Though we concern ourselves in the case $M >m$, it is 
meaningful to investigate the degree of ambiguity near the equal 
mass limit. 
For $M=m$, the projection factor in front of the interaction
 term of (\ref{4}) includes a part which violates the exchange 
symmetry: 
It is shown that
\begin{equation}
  \Lambda^{(V)}=\Lambda_{+-}-\Lambda_{-+}         \label{e1}
\end{equation}
and the Heisenberg's exchange operator
\begin{equation}
  P_H=\frac{1}{4}(1+\bmth{\sigma}_1\cdot\bmth{\sigma}_2)
   (1+\bmth{\rho}_1\cdot\bmth{\rho}_2)P_M              \label{e2}
\end{equation}
anticommute, where the operator $P_M$ exchanges the
 momenta(or coordinates). $\Lambda^{(V)}$ violates the symmetry since
 the remaining part of the Hamiltonian and $P_H$ commute. 

For equal-mass limit, we have two equations. One is the equation 
(\ref{4}) with $M=m$ and another is obtained by assigning a minus sign
 in front of $\Lambda^{(V)}$,
 which is the equal-mass limit of the equation with
$m>M$. It is the conjugate equation of (\ref{4}) in the sense
 that (\ref{4}) is converted into it by the transformation $P_H$.
 It is easy to show that these equations have the common eigenvalue
spectrum:
 If an eigenfunction $\chi_n$ of (\ref{4}) belongs to some eigenvalue
$E_n$, $P_H\chi_n$ is the solution of the conjugate equation with the
 same eigenvalue. However, $\chi_n$ does not have the definite 
$P_H$-parity.

For equal masses, it is reasonable to use the Salpeter equation 
which includes the projection factor (\ref{2}). 
The eigenvalue of this equation is different from the above $E_n$.
 The difference is, however, small since it is of the 4th order
in the symmetry-breaking part of the Hamiltonian.

\section{Summary}

By assuming a proper boundary condition for the two-body propagation 
in the negative-energy state, we have proposed a new 
bound-state equation for the unequal-mass constituents. 
It has the symmetrical energy eigenvalues $E$ and $-E$ and reduces 
to the (one-body)Dirac equation in the infinite limit of one of the 
constituents masses. Secondly, we have discussed the normalization of
 the wave function and pointed out that the positive-definite 
probability density should be assumed.
 We can consistently calculate the observables of the bound state by 
assuming this normalization.

The interaction Hamiltonian of the equation is constructed by 
diagonalizing the field theoretic Hamiltonian in the generalized Fock 
subspace of the two particles. Relativistic effects such as the 
retardation are taken into account in systematic way in the framework 
of the perturbation theory.

We have further investigated the WFO's of the proposed equation
 in some detail for the instantaneous interaction and shown that the 
average WFO in the ${}^1S_0$ state which determines the leptonic 
decay rate of a pseudo-scalar meson is finite, which is in accord 
with the expectation from the field theory.

One of the system for which we may utilize the two-body Dirac 
equation is the heavy flavored quarkonium. There are various 
possibilities of choosing the framework of the approximation 
according to the treatment for the high-momentum interaction. We have 
briefly discussed the correction factor to the leptonic decay width 
caused by the high momenta. The equation affords a good foundation 
of the heavy quark effective theory.

 \section*{Acknowledgements}

The author would like to thank Professor G.A. Kozlov for useful
 communication on the two-time Green's function.

\subsection*{Appendix }

We examine the asymptotic
($p\rightarrow\infty$) behavior of the momentum-space wave function
 and show that the average WFO 
$\mbox{Tr}\{ \gamma_5\gamma_0\psi(0) \}/\sqrt{2}$
 is finite.\footnote{See Ref.\cite{Ito88} and references therein, for
 the Salpeter equation.}


There are 4 partial amplitudes $h_{\epsl\eta}(p)$ in the ${}^1S_0$
 state,
with which the wave function is expanded as

\begin{equation}
\chi(\bmth{p})=\sum_{\epsl\eta}\sum_{r}c_ru_\epsl^r(-\bmth{p})
v_\eta^{-r}(\bmth{p})h_{\epsl\eta}(p)
( \frac{1}{16\pi E_1E_2} )^{1/2},
                                                      \label{pw}
\end{equation}
where $c_{1/2}=-c_{-1/2}=1/\sqrt{2}$ and the spinors $u$ and $v$, for
 the particle 1 and 2 respectively, are defined in \cite{Ito82}.

The average WFO for the annihilation decay through the axial-vector
 current is given by

\begin{eqnarray}
\lefteqn{ \frac{1}{\sqrt{2}}\mbox{Tr}\{ \gamma_5\gamma_0\psi(0) \}
 = \frac{1}{\sqrt{8}\pi}\int (\frac{1}{E_1E_2})^{1/2} }     \nonumber
 \\ & & \times\sum_{\epsl\eta}
\{ \sqrt{(E_1+\epsl M) (E_2+\eta m)}-\epsl\eta
\sqrt{ (E_1-\epsl M)
 (E_2-\eta m) } \}h_{\epsl\eta}(p)p^2dp. \hspace{6mm}   \label{wf}
\end{eqnarray}

We first assume the Coulomb potential. The partial-wave equation for
 the ${}^1S_0$ state is given by

\begin{eqnarray}
 \lefteqn{ \{ E-\epsl E_1(p)-\eta E_2(p) \}h_{\epsl\eta}(p)
=-\epsl\frac{\alpha}{4\pi}\sum_{\epsl'\eta'}\int dq } \nonumber \\
 & & \times \frac{q}{p}[ \frac{1}{E_1(p)E_1(q)E_2(p)E_2(q)} ]^{1/2}
 [\{ A^1_{\epsl\epsl'}A^2_{\eta\eta'}
+\epsl\epsl'\eta\eta'A^1_{-\epsl-\epsl'}A^2_{-\eta-\eta'} \}Q_0(z)
 \nonumber \\
 & &  \mbox{} +\{\epsl\epsl' A^1_{-\epsl-\epsl'}A^2_{\eta\eta'} +
 \eta\eta'A^1_{\epsl\epsl'}A^2_{-\eta-\eta'} \}Q_1(z)]
   h_{\epsl'\eta'}(q),                      \label{pwe}
 \end{eqnarray}
where $z=(p^2+q^2)/2pq$ and $Q_\ell(z)$ is the Legendre's function.
$A^i_{\epsl\epsl'}$ is defined by

\[ A^i_{\epsl\epsl'}=\sqrt{(E_i(p)+\epsl m_i)(E_i(q)+\epsl' m_i)}. \]

 The asymptotic behavior of the wave function is determined from the 
integral region near the infinity.
 We then expand the both sides of (\ref{pwe}) into the series of
 $1/p$ and $1/q$. 
We assume the power behavior of the wave function for large $p$.
The independent amplitudes are chosen to be
$ h_A(p)\equiv\sum_\epsl h_{\epsl\epsl}(p)$,
$h_B(p)\equiv\sum_\epsl\epsl h_{\epsl\epsl}(p)$,
$h_C(p)\equiv\sum_\epsl h_{\epsl-\epsl}(p)$,
and
$h_D(p)\equiv\sum_\epsl\epsl h_{\epsl-\epsl}(p)$,
which are expanded, in the high-momentum region, in power series of
 $1/p$:
\[ h_X(p)=\sum_n C_X^np^{-\beta_X-2n-1}. \]

Integrals on the right-hand side can be done if we neglect infrared-
divergent terms which are irrelevant to the leading asymptotic
 behavior. Now, we can determine the asymptotic indices
 $\beta_X$'s from consistency\cite{Ito82}:
We get, for $h_A$ and $h_B$,

\begin{equation}
 2C_A^0 p^{-\beta_A}- EC_B^0 p^{-\beta_B-1}=
  \frac{\alpha}{\pi}C_A^0 \frac{\pi}{1-\beta_A}\cot
(\frac{\pi}{2}\beta_A)
p^{-\beta_A}    \label{ia}
\end{equation}
and
\begin{equation}
 2C_B^0 p^{-\beta_B}- EC_A^0 p^{-\beta_A-1}=
  \frac{\alpha}{\pi}C_B^0
\frac{\pi(1-\beta_B)}{\beta_B(2-\beta_B)}\tan(\frac{\pi}{2}\beta_B)
  p^{-\beta_B},    \label{ib}
\end{equation}
where the terms of the higher power in $1/p$ are neglected.
If we neglect the second term in the left-hand sides of (\ref{ia}),
we find $\beta_A$ in the range $1<\beta_A<2$
\footnote{See Ref.\cite{Ito82} for the details.}
 and get $\beta_B=\beta_A+1$ from (\ref{ib}). We obtain another
 series by neglecting the second term in (\ref{ib}). For this,
 $\beta_B$ is found to be in the range $2<\beta_0<\beta_B<3$,
where the lower bound $\beta_0$ corresponds to the
upper bound $4/\pi$ of $\alpha$ above which the index $\beta_A$ from
(\ref{ia}) becomes complex.
$\beta_A$ of the second series is given by $\beta_A=\beta_B+1$.

The asymptotic amplitudes $h_C$ and $h_D$ are determined dependently
 on $h_A$ and $h_B$.
We get, for the minimum indices

\begin{equation}
 \beta_C=\min(\beta_A+2,\: \beta_B+1)
\end{equation}
\begin{equation}
 \beta_D= \beta_A+1.
\end{equation}
We see that
the average WFO (\ref{wf}) is finite, because
\[  \beta_B>1 \: \mbox{ and }\: \beta_C>2               \]
hold for the asymptotic amplitudes.
This conclusion is valid even if the
 instantaneous exchange(transverse part) of the gauge boson is added.

\nocite{Feynman49}
\nocite{Hayashi52}
\nocite{Breit30}
\nocite{Ito901}
\nocite{Ito93}
\nocite{Aitchison82}

\end{document}